\documentclass[aps,preprint,floats,prb]{revtex4}
\usepackage{graphics,dcolumn}

\begin{document}

\title{Stacking of oligo and polythiophenes cations in solution:
surface tension and dielectric saturation}
\author{Dami\'{a}n A. Scherlis$^{\dag,\ddag}$}
\author{Jean-Luc Fattebert$^\S$}
\author{Nicola Marzari$^{\ddag}$}
\affiliation{$^\dag$Departamento de Qu\'{i}mica Inorg\'{a}nica, Anal\'{i}tica
y Qu\'{i}mica F\'{i}sica, Facultad de Ciencias Exactas
y Naturales, Universidad de Buenos Aires, Ciudad Universitaria,
Pab. II, Buenos Aires (C1428EHA) Argentina}
\affiliation{$^\ddag$Department of Materials Science and Engineering,
and Institute for Soldier Nanotechnologies,
Massachusetts Institute of Technology, Cambridge MA 02139}
\affiliation{$^\S$Center for Applied Scientific Computing, Lawrence
Livermore National Laboratory, Livermore CA 94551}

\begin{abstract}
 
The stacking of positively charged (or doped) terthiophene oligomers and
quaterthiophene polymers in solution is investigated applying a recently
developed unified electrostatic and cavitation model for first-principles
calculations in a continuum solvent. The thermodynamic and structural
patterns of the dimerization are explored in different solvents,
and the distinctive roles of polarity and surface tension are
characterized and analyzed. Interestingly, we discover a saturation
in the stabilization effect of the dielectric screening that takes place
at rather small values of $\epsilon_0$. Moreover, we address the interactions
in trimers of terthiophene cations, with the aim of generalizing the
results obtained for the dimers to the case of higher-order stacks and
nanoaggregates.

\end{abstract}

\date{\today}
\maketitle

\section{Introduction}

The intermolecular interactions between
conjugated polymers and oligomers in the condensed phase---whether
in the solid state or in solution---entail a fundamental interest
in the emerging field of molecular electronics,
as they are decisive factors in the
electronic and structural properties of these materials.
The orientation and alignment of polymers in a solid matrix,
the formation of aggregates in films and solution,
or the ability of organic semiconductors to self-assemble are the outcome
of a complex balance between the spatial features of the molecules
and the substrate, and the interactions
between them at the conditions of synthesis.\cite{nguyen-sci}$^-$
\cite{sirringhaus}
By dictating rules for aggregation, these interactions
eventually shape properties such as charge
delocalization and mobility\cite{schwartz,sirringhaus}$^-$\cite{bredas1}
or optical\cite{nguyen-jpc,bredas}$^-$\cite{mishina} and
electromechanical\cite{smela} response.

Thiophene-derived oligomers and polymers represent today one
of the most promising class of organic semiconductors, finding
potential applications in a variety of electronic and electroactive
devices.\cite{gross}$^-$\cite{ourjacs} Semiconducting properties
arise with doping, therefore much of the basic research performed on these
systems has addressed in particular the doped or oxidized species.
Since the early nineties electrochemical and spectroscopic evidence 
was gathered indicating that oxidized oligothiophenes reversibly associate 
in solution.\cite{miller92}$^-$\cite{roncali} In a recent
study,\cite{ourjpc} we have outlined how this association
is driven by three contributions: the 
attractive $\pi$-$\pi$ interactions, the Coulombic repulsion,
and the solvent effects. In the case of oligothiophene
cations dimers, combination of
semioccupied HOMOs form occupied bonding and empty antibonding
orbitals, resulting in an interaction of covalent character, different
in nature to the one arising in neutral dimers, of
dispersive origin.\cite{suzuki}
In vacuum, the electrostatic
repulsion between the cations largely exceeds the covalent term,
\cite{ourjpc, brocks}
making apparent the importance of the solvent (or of the counterions
in the solid state case) in stabilizing the stacks. A polarizable
dielectric medium favors concentration of charge in a small cavity,
reverting the balance from net repulsion to attraction and stacking.

In the present paper, we employ a recently developed first-principles
approach recently developed 
to describe the effect of a continuum solvent within the 
density-functional theory framework,\cite{solvation}
and use it to explore the role of polarity
and surface tension in the stabilization of dimers of
polythiophene and oligothiophene radical cations.
Furthermore, we examine the possibility of trimer formation 
as an intermediate step toward the nucleation of higher-order aggregates,
and to gain insight on self-assembly in solution.
In charged dimers, at variance with the case of neutral dimers governed by
van-der-Waals forces, interactions are
predominantly covalent and electrostatic, and density-functional theory
(DFT) has proven to be sufficiently accurate when compared
with highly-correlated
quantum-chemistry methods.\cite{ourjpc}
In our approach,
the contribution of the surface tension to the solvation free
energy is computed in a very natural fashion, as the product 
between the area of the cavity and
the surface tension of the solvent.\cite{solvation}
This contribution is particularly important in dimerization processes,
where the merging of two cavities into one provides an additional
stabilizing term associated to the minimization 
of the total area of the cavity.

\section{Methodology}

All calculations in this work have been performed with the
public domain Car-Parrinello parallel code included in the Quantum-ESPRESSO
package,\cite{Espresso} based on density-functional theory,
periodic-boundary conditions, and plane-wave basis sets.
Vanderbilt ultrasoft pseudopotentials\cite{usp} have been used
to represent the ion-electron interactions, in combination with the
PBE approximation to the exchange-correlation term,\cite{pbe} and with
Kohn-Sham orbitals and charge density expanded in plane
waves up to a kinetic energy cutoff of 25 and 200 Ry respectively.

Solvation is described with a continuum model recently implemented
by us in the
Quantum-ESPRESSO package and described
in detail in reference ~\cite{solvation}.
In this approach, the solvent is represented as
a dielectric medium surrounding a quantum-mechanical
solute confined in a cavity delimited by an isosurface
of electronic charge density. Adopting a common decomposition of
the solvation free energy
$\Delta G_{sol}$ we have:
\begin{equation}
\Delta G_{sol} = \Delta G_{el}+\Delta G_{cav}+
\Delta G_{dis-rep}
\end{equation}
where $\Delta G_{el}$, $\Delta G_{cav}$, and $\Delta G_{dis-rep}$
are the electrostatic, the cavitation, and
the dispersion-repulsion contributions respectively.\cite{crtomasi1}
In our implementation
$\Delta G_{el}$ and $\Delta G_{cav}$ are considered explicitly,
while $\Delta G_{dis-rep}$
is largely captured (by virtue of the
parametrization) by the electrostatic term. In the following, we briefly review
the approaches used to obtain $\Delta G_{el}$ and $\Delta G_{cav}$.

The electrostatic interaction between the dielectric medium and the solute
is calculated, as proposed by Fattebert and Gygi,\cite{jluc2,jluc1} by
solving the Poisson equation in the presence of a dielectric
continuum with permittivity $\epsilon[\rho]$:
\begin{equation}
\nabla \cdot (\epsilon[\rho] \nabla \phi) = -4\pi \rho~.
\end{equation}
The function $\epsilon[\rho]$ is defined to
asymptotically approach the permittivity of the bulk solvent $\epsilon_0$
in regions of space where the
electron density is low, and to approach 1 in those
regions where it is high.\cite{solvation}
In this way the dielectric medium and the
electronic density respond self-consistently to each other
through the dependence of $\epsilon$ on $\rho$ and vice-versa.
The variation in the dielectric constant at the solvent-solute
interface is controlled by two parameters $\rho_0$ and
$\beta$, which determine the size of the cavity and the smoothness
of the transition region. These are the only parameters
entering the model, and our chosen values, $\rho_0$=0.00078 $e$ and
$\beta$=1.3, represent a rather universal choice.\cite{solvation}

The cavitation term is computed as the product between the surface
tension of the solvent $\gamma$ and the area of the cavity,
\begin{equation}
\Delta G_{cav} = \gamma S(\rho_0),
\end{equation}
where $S(\rho_0)$ is the surface of the same cavity employed in the
electrostatic part of the solvation energy and is defined by an isosurface
of the charge density. 
This area can be easily and accurately calculated by
integration in a real-space grid, as the volume of a thin film delimited
between two charge density isosurfaces, divided by the thickness of this film.
This idea has been originally proposed by Cococcioni et al.\cite{matteo}
to define a ``quantum surface'' in the context of extended electronic-enthalpy
functionals:
\begin{equation}
S(\rho_0)= \int d{\bf r} \left\{ \vartheta_{\rho_0-
\frac{\Delta}{2}} \left[ \rho({\bf r})\right]
-\vartheta_{\rho_0+\frac{\Delta}{2}} \left[ \rho({\bf r})
\right] \right\} \times \frac{|\nabla \rho ({\bf r})|}{\Delta}.
\end{equation}
The finite-differences parameter $\Delta$ determines the separation between two
adjacent isosurfaces, one external and one internal, corresponding to
density thresholds $\rho_0 - \Delta/2$ and $\rho_0 + \Delta/2$
respectively. The spatial distance between these two cavities---or the
thickness of the film---is given at any point in space by
the ratio $\Delta/|\nabla \rho|$.
The (smoothed) step function $\vartheta$ is zero in regions of low
electron density and approaches 1 otherwise, and it has been defined
consistently with the dielectric function $\epsilon[\rho]$.

\section{Results and discussion}

Crystallographic data\cite{graf} and recent
calculations\cite{ourjpc,brocks} on oxidized dimers have indicated
that the stacking of oligothiophene cations follows a ``slipped''
pattern where the layers are shifted along the molecular axis, as shown
in Fig.~\ref{terthdimer}. We have studied the dependence of the
energy as a function of the lateral shift for oxidized
terthiophene and polyquaterthiophene dimers in acetonitrile
($\epsilon_0$=35.7 and $\gamma$=28.7 mN/m) at a fixed
intradimer separation of 3.4 $\AA$. In the calculations involving the
terthiophene oligomers, Dirichlet boundary conditions in
the electrostatic potential were applied, and the size 
of the unit cell was large enough to eliminate any significant
interactions between the periodic images. In the case of the polymers,
periodic boundary conditions were used along the {\it z} axis (the molecular
axis), keeping the Dirichlet boundary conditions in the other directions.
The total charge of these systems is +2 (in the
polymer, there is a positive charge every four thiophene rings). 
Our results are shown
in Fig.~\ref{displ}: both the oligomer and the polymer exhibit a similar
pattern, with a global minimum at 2.3 $\AA$ for the terthiophene
and 2.0 $\AA$ for the polyquaterthiophene. A local minimum at 0.0 $\AA$
(where the two layers are overlapping) is present
in both cases. Interestingly, the net binding is
very sensitive to the lateral shift, varying steeply in a range of
10 kcal/mol as one layer is slipped over the other. At shifts
of about 1 $\AA$ off the minima, the $\pi$-$\pi$ interaction
between the cations appears clearly weakened, resulting
in an unbound dimer. 

The nature of the solvent doesn't have any
significant effect on this characteristic pattern, even if it
affects the magnitude of the interaction. This is shown
in Fig.~\ref{displs}, where the terthiophene curve is displayed
for three different media: acetonitrile,
dichloromethane ($\epsilon_0$=8.9, $\gamma$=27.2 mN/m) and water 
($\epsilon_0$=78.8, $\gamma$=72.2 mN/m).(We have chosen water as a
case study given its distinctive polarity and surface tension,
despite the low solubility exhibited by thiophene derivatives in this solvent.)
Fig.~\ref{evsdt} explicitly illustrates the role of the solvent in the
binding of the terthiophene cations, by showing the interaction energy
as a function of the intradimer distance at a
fixed lateral shift of 2.3 $\AA$. The binding energies are close
to 5 kcal/mol for dichloromethane and acetonitrile, and 12 kcal/mol
in water. These energies can be seen as a lower limit for the
absolute value of the dimerization enthalpy $\Delta H_d$,
since the effect of the ionic
environment was neglected in the calculations
(the counterions in solution would differentially stabilize
the doubly-charged terthiophene dimer compared to two terthiophene
cations. This effect has been recently discussed by Jakowski and 
Simons\cite{simons} for dimers of tetracyanoethylene anions,
[TCNE]$_2^{2-}$). In fact,
dimerization enthalpies between 7 and 14 kcal/mol
have been reported for different terthiophene derivatives in apolar
solvents.\cite{bauerle}$^-$\cite{hapiot}
The interplanar separations corresponding to the minima, in the range
of 3.4 to 3.5 $\AA$, are consistent with the distance of 3.47 $\AA$
obtained for substituted terthiophene cations in the solid state.\cite{graf}
Fig.~\ref{evsdp} compares the potential energy surface of the
oligomer with the one corresponding to the polymer
(the later was calculated at a fixed lateral shift of 2.0 $\AA$).
The equilibrium distance turns out around
0.3 $\AA$ larger in the periodic system, although it exhibits
a slightly stronger binding. This is
in agreement with experimental data showing 
that $\Delta H_d$ is enhanced by the length of the chain,
\cite{miller-acc} a trend related to a ``dilution'' of
the Coulombic repulsion as the ratio between charge and
oligomer size decreases.\cite{ourjpc} At the same time,
however, the increase in length at a given oxidation state
would diminish the ratio between unpaired electrons
available to $\pi$-$\pi$ bonding and thiophene rings, what
would presumably revert the aforementioned binding trend
starting from certain molecular weighs.\cite{nota1}

The separate roles played by the dielectric screening of the
solvent and its surface tension in the stabilization of the dimer are
highlighted in Fig.~\ref{cav}. If the contribution of $\Delta G_{cav}$
were omitted, the binding curves would turn out to be
very close to each other (Fig.~\ref{cav}a).
The larger $\gamma$ in the case of water (72.2 mN/m versus
28.7 mN/m in acetonitrile) is responsible for the deeper minimum in
the potential energy surface. The net effect of the surface tension
is to minimize the area of the solvation cage, monitored in
Fig.~\ref{cav}b as the cations are pulled apart. Beyond a
separation of about 4.75 $\AA$ the surface remains constant,
indicating that the single cavity has split and each cation is enclosed 
in a separate cavity of area independent of the interplanar distance.
In this situation there is no cavitation energy gain
and therefore the curves excluding and including $\Delta G_{cav}$
(open and closed symbols in Fig.~\ref{cav}a respectively)
overlap on the right part of the plot.

The potential energy curves in Fig.~\ref{cav}a unveil an intriguing 
possibility: that the binding energy is not directly related to the dielectric
constant of the solvent, as our intuition may suggest. This hypothesis
is explored in Fig.~\ref{epsilon}, where the interaction energy
between two terthiophene cations separated by 3.6 $\AA$
is plotted as a function of the dielectric constant, ignoring the
contribution of the surface tension. The results are somehow
unexpected: a rather small increase in the permittivity with respect to
the vacuum limit rapidly stabilizes the dimer, but once the dielectric 
constant is above 10 the effect of a further increase in polarity is very small.
This behavior can be rationalized considering that a polarizable dielectric
medium with low permittivity is already enough to 
screen most of the Coulombic repulsion between the two charges
and to favor aggregation of these charges by polarizing itself.
We note in passing that
the positive drift observed at higher permittivities
for the case of $\rho_0$=0.00078 $e$
is an artifact of the continuum model. Since the dielectric constant
is defined as a continuous function of the electron density, its value
throughout the intradimer region may depart from 1, allowing the
dielectric medium to fill some of the space between the cations and
to interfere, though modestly, with the $\pi$-$\pi$ bond. This effect
will be enhanced at large values of $\epsilon_0$ and $\rho_0$.
In reality, instead, the solvent does not penetrate the 
intradimer space if the separation is 3.6 $\AA$, regardless 
of $\epsilon_0$. This spurious behavior is in fact absent in the curve 
computed with $\rho_0$=0.0003 in Fig.~\ref{epsilon}.
What is remarkably captured 
by the continuum model is the saturation
effects of polarity on the dimerization, occurring already for very
low dielectric constants.
These results are pretty much consistent with
experimental observations that turn down a direct correlation
between dimerization trends and polarity of the 
medium, while emphasizing the dependence on solubility
of the oligothiophenes.\cite{janssen} To understand the effect of the
solvent on $\Delta H_d$, then, one should consider other properties
such as surface tension or specific interactions between the
solute and the medium.

Are the thermodynamic and structural features found so far for the
dimerization applicable to the stacking of multiple oligomer layers?
It would be very interesting to know if or how the present results can be
extended to processes such as aggregation and self-assembly in solution,
involving the collective pairing of many oligothiophene units.
In an attempt to offer an answer, even if preliminary, to this question,
we have studied the formation of trimers of terthiophene cations 
in acetonitrile. Fig.~\ref{trimer} depicts the two configurations 
of minimum energy obtained for the trimer in acetonitrile, in which the
third cation is shifted by + or - 2.3 $\AA$ with respect to
the next oligomer. As shown in Fig.~\ref{trimerevsl}, where
the interaction energy is plotted as a function of the lateral shift
of the third cation, there is no
significant energetic difference between these two minima.
The curve corresponding to the dimer
is plotted in the same figure: the pattern of valleys and
peaks is preserved at the same lateral displacements 
when increasing the number of layers from two
to three. The differences in the relative depths of these curves
can be ascribed to the fact that the same interplanar separation
of 3.4 $\AA$ was adopted in the calculation of both,
but the optimal separation in the trimer is
longer, as can be seen in Fig.~\ref{trimerevsd}. This graph
shows the interaction energy calculated
for the trimer in acetonitrile
as a function of the interplanar separation between layers
(the interplanar separation between the first
and the second layer is the same as between the second and the third
at each point of the curve). For meaningful
comparison with the dimer, depicted in the same graph, the
energies were normalized to the number of $\pi$-pairs,
in this case two. Interestingly, the binding
between two cations doesn't seem to be impaired by the presence
of a third one: the interaction between stacks remains
almost constant, even though the equilibrium distance increases 
in about 0.1 $\AA$. This suggests that the energetic and structural
results found for the cation dimers can be applied,
to a large extent, to the case of more complex, larger aggregates
consisting of multiple layers.

\section{Summary}

Our study has highlighted the separate
roles of surface tension
the dielectric screening in the stabilization of charged thiophene oligomers
and polymers stacks. The surface tension of the solvent is a driving
force toward the minimization of the cavity area, and therefore toward
dimerization: there is an energetic payoff
in accommodating two solutes in a single
cavity of an area smaller than twice the one corresponding to the
dissociated components. On the other hand, the dependence of
the dimer stability on the polarity of the solvent alone is less
evident. A dielectric effect is necessary to screen the
electrostatic repulsion and to stabilize the charges in a small
volume, but once the permittivity has reached a certain threshold,
a further increase in polarity has a negligible contribution to 
the stabilization of the system. This observation is probably general
to any $\pi$-dimer of charged radicals---an hypothesis that could
be interesting to test through explicit calculation.

The formation of trimers follows the same geometrical arrangement as
the dimerization. A $\pi$-bond on one of the oligomer planes
does not seem to significantly affect the bond on the other. These
results point to the conclusion that the organization of aggregates and
stacks is governed by the same thermodynamics that is already manifest
in the dimerization.

\section{Acknowledgments}
This research was supported by the MURI Grant DAAD 19-03-1-0169,
by the Institute of Soldier
Nanotechnologies, contract DAAD-19-02-D0002, and by
Fundaci\'{o}n Antorchas.
A portion of this work was performed under the auspices of the U.S.
Department of Energy by University of California Lawrence Livermore 
National Laboratory under contract No. W-7405-Eng-48.

\newpage

\noindent{\bf Figure Captions:}\\

\noindent{\bf Figure 1.}:
A terthiophene dimer in the minimum energy configuration corresponding
to the doubly-charged state, in
which one singly-charged monomer is shifted 2.3 $\AA$ with respect 
to the other along the main axis.

\noindent{\bf Figure 2.}:
Interaction energy as a function of the axial shift for the oxidized
terthiophene and quaterthiophene dimers in acetonitrile.

\noindent{\bf Figure 3.}:
Interaction energy as a function of the axial shift for the
oxidized terthiophene dimer in water ($\epsilon_0$=78.8, $\gamma$=72.2 mN/m),
acetonitrile ($\epsilon_0$=35.7, $\gamma$=28.7 mN/m), and
dichloromethane ($\epsilon_0$=8.9, $\gamma$=27.2 mN/m).

\noindent{\bf Figure 4.}:
Interaction energy as a function of the interplanar separation
between two singly-charged terthiophene cations
in water, acetonitrile, and dichloromethane.

\noindent{\bf Figure 5.}:
Interaction energy as a function of the interplanar separation
for the doubly-charged dimers of polyquaterthiophene (open symbols)
and terthiophene (closed symbols) in dichloromethane and acetonitrile.

\noindent{\bf Figure 6.}:
(a) Interaction energy of two terthiophene cations in
acetonitrile and in water, as a function of its separation.
Open symbols curves were calculated omitting
the cavitation contribution to the solvation energy,
while the closed symbols curves include both the electrostatic and
cavitation contributions.
(b) Area of the solvation cavity as a function of the
separation between the terthiophene cations. Beyond 4.75 $\AA$
the cavity splits in two, and the plotted area corresponds to
two cavities containing one singly-charged terthiophene each.

\noindent{\bf Figure 7.}:
Interaction energy between two terthiophene cations at a fixed
separation as a function of the dielectric constant of the
solvent, omitting the cavitation energy term, for two different
thresholds $\rho_0$. The positive drift observed at high permittivity
for $\rho_0$=0.00078 $e$ is an artifact of the continuum model (see text).

\noindent{\bf Figure 8.}:
Minimum energy structures for the terthiophene trimer in acetonitrile,
surrounded by its corresponding solvation cavities
defined by isosurfaces at 0.00078$e$.

\noindent{\bf Figure 9.}:
Interaction energy as a function of the axial shift between a 
doubly-charged terthiophene 
dimer and a third terthiophene cation in acetonitrile.

\noindent{\bf Figure 10.}:
Interaction energy as a function of the interplanar separation between
three parallel terthiophene cations in acetonitrile. The top and the
bottom layers are overlapping with each other,
having an axial shift of 2.3 $\AA$ with
respect to the central layer, as shown in Fig.~\ref{trimer}a.

\newpage

\begin{figure} 
\caption{D. Scherlis}
\label{terthdimer}
\end{figure}

\newpage

\begin{figure} \centerline{
\rotatebox{-90}{\resizebox{5in}{!}{\includegraphics{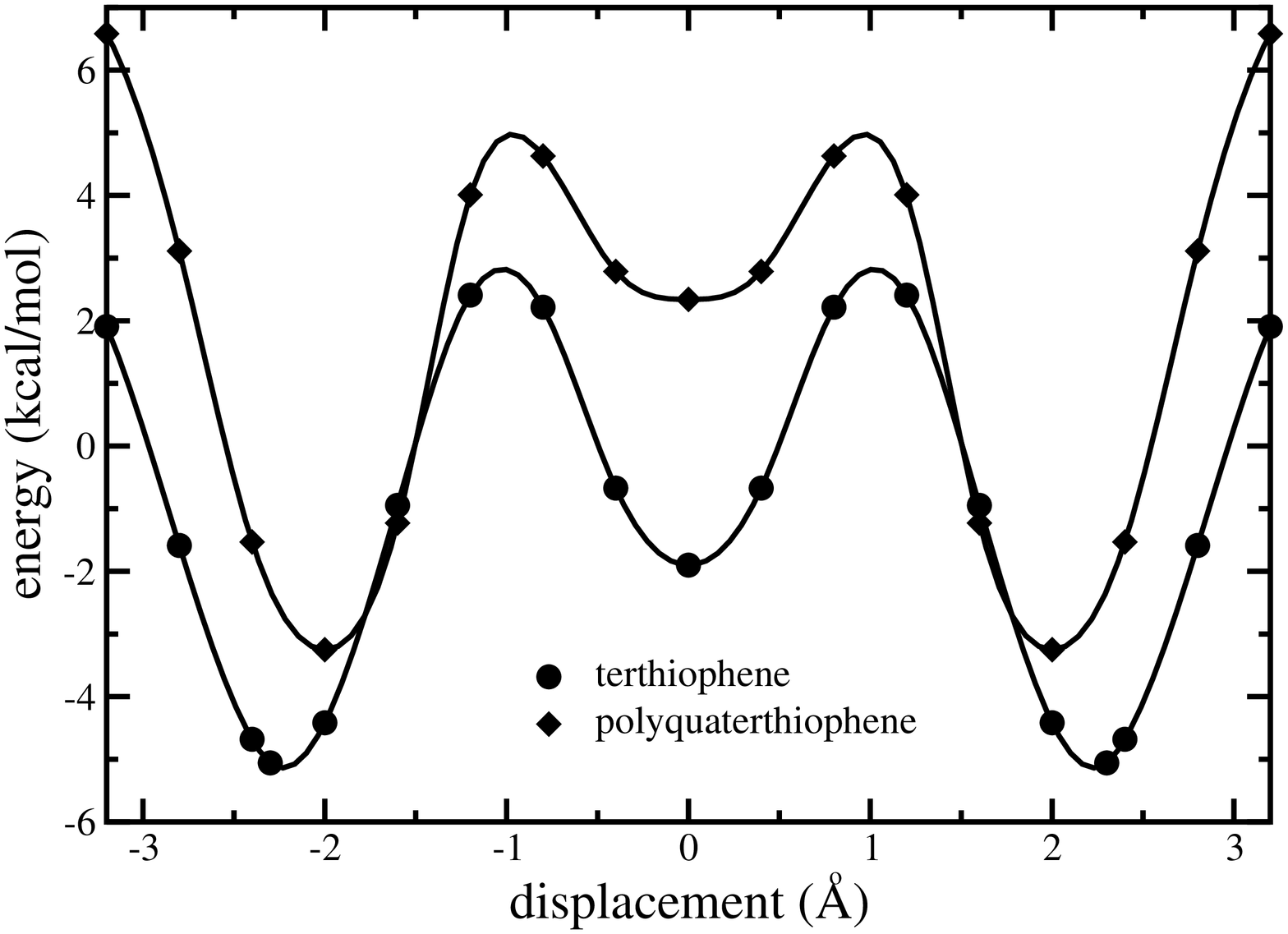}}}
}
\caption{D. Scherlis}
\label{displ}
\end{figure}

\newpage

\begin{figure} \centerline{
\rotatebox{-90}{\resizebox{5in}{!}{\includegraphics{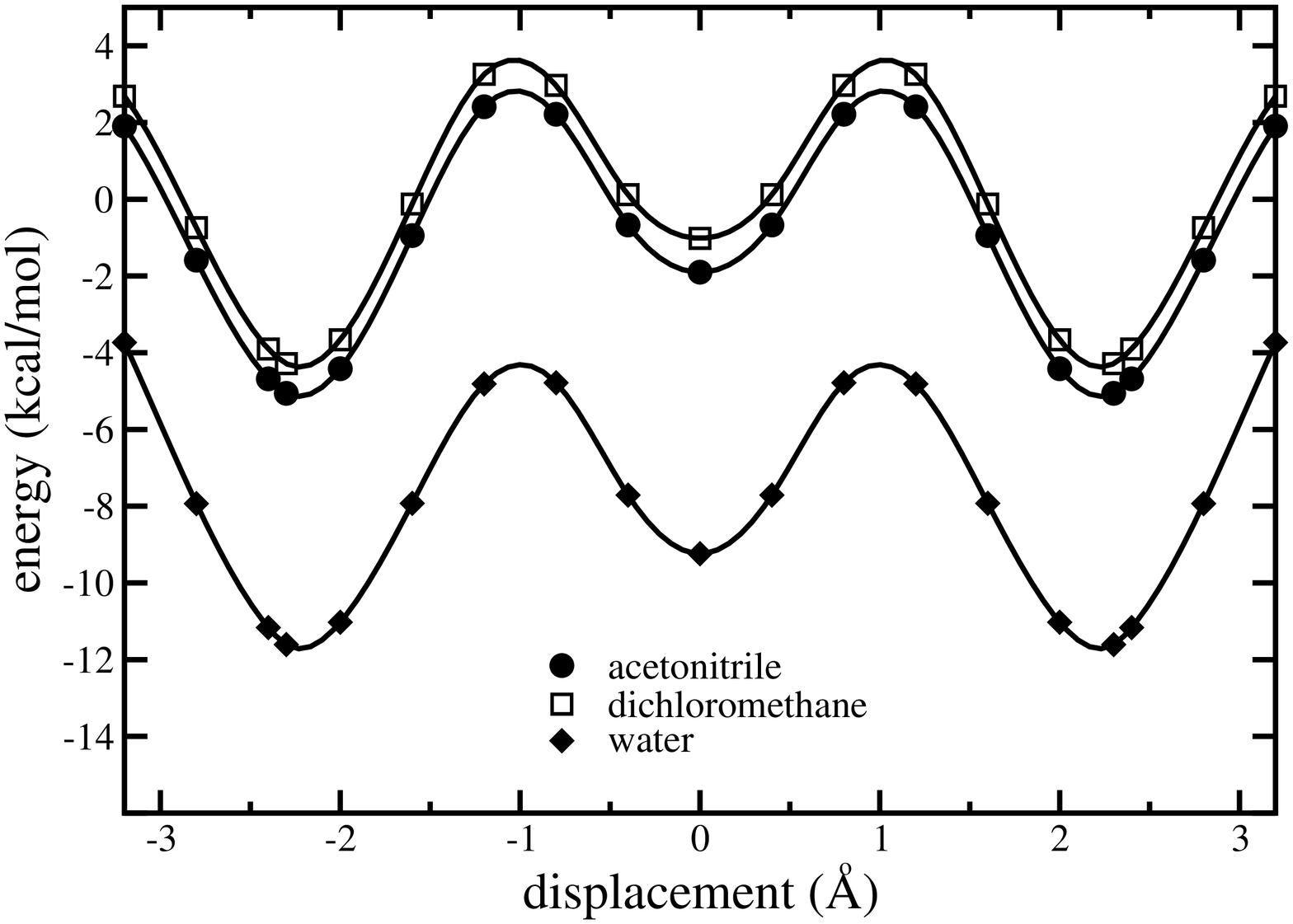}}}
}
\caption{D. Scherlis}
\label{displs}
\end{figure}

\newpage

\begin{figure} \centerline{
\rotatebox{-90}{\resizebox{5in}{!}{\includegraphics{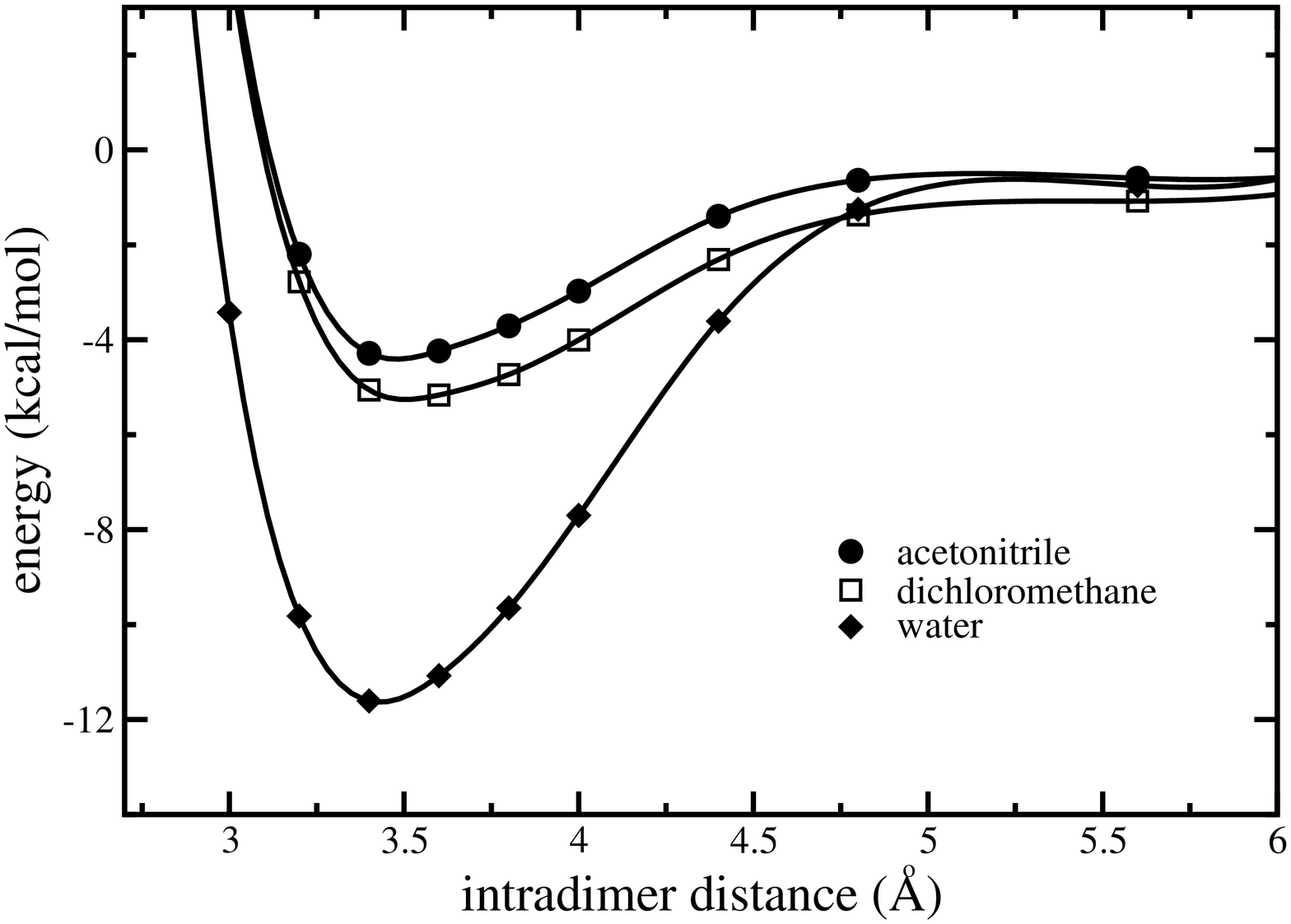}}}
}
\caption{D. Scherlis}
\label{evsdt}
\end{figure}

\newpage

\begin{figure} \centerline{
\rotatebox{-90}{\resizebox{5in}{!}{\includegraphics{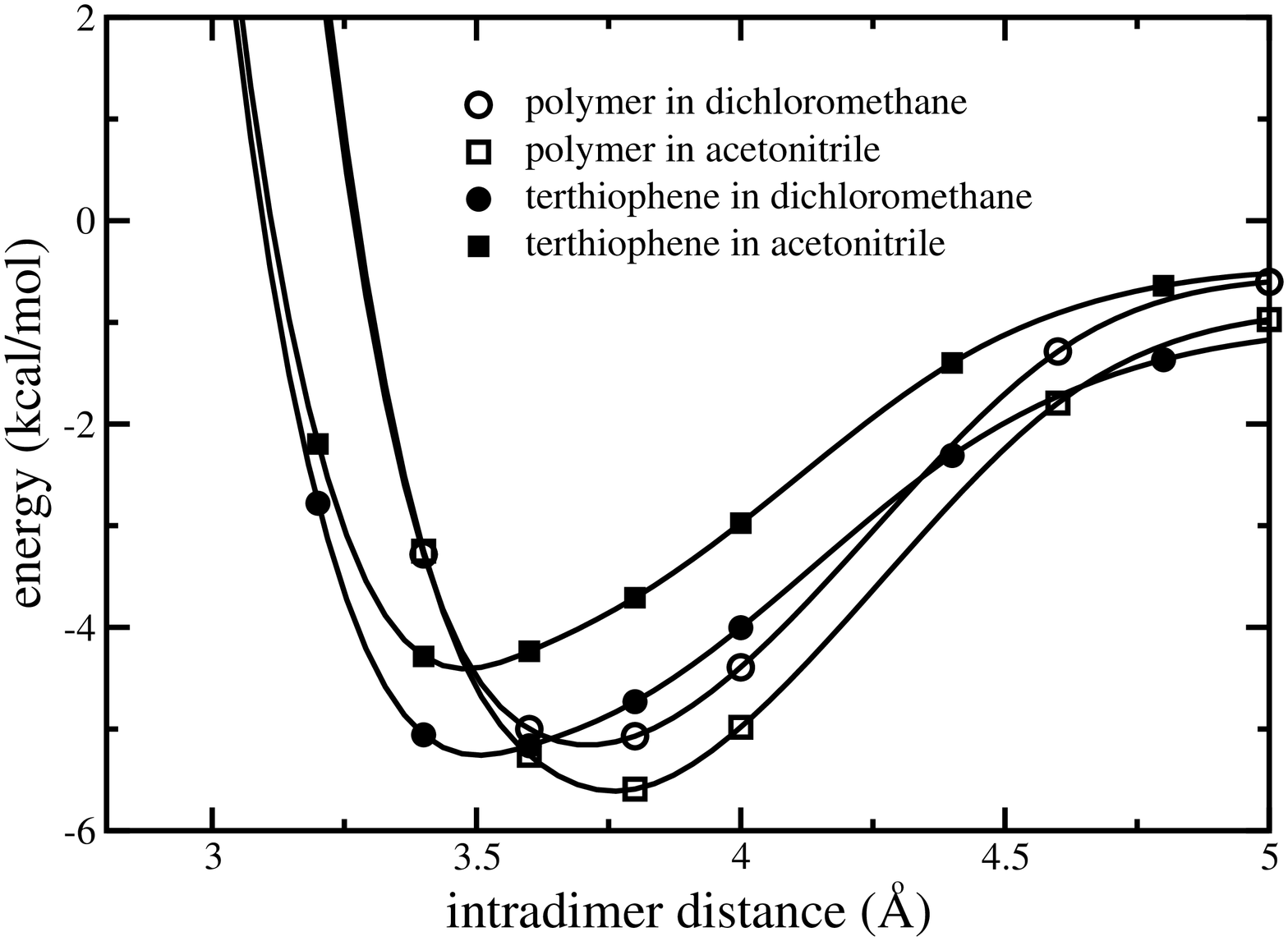}}}
}
\caption{D. Scherlis}
\label{evsdp}
\end{figure}

\newpage

\begin{figure} \centerline{
\rotatebox{-90}{\resizebox{5in}{!}{\includegraphics{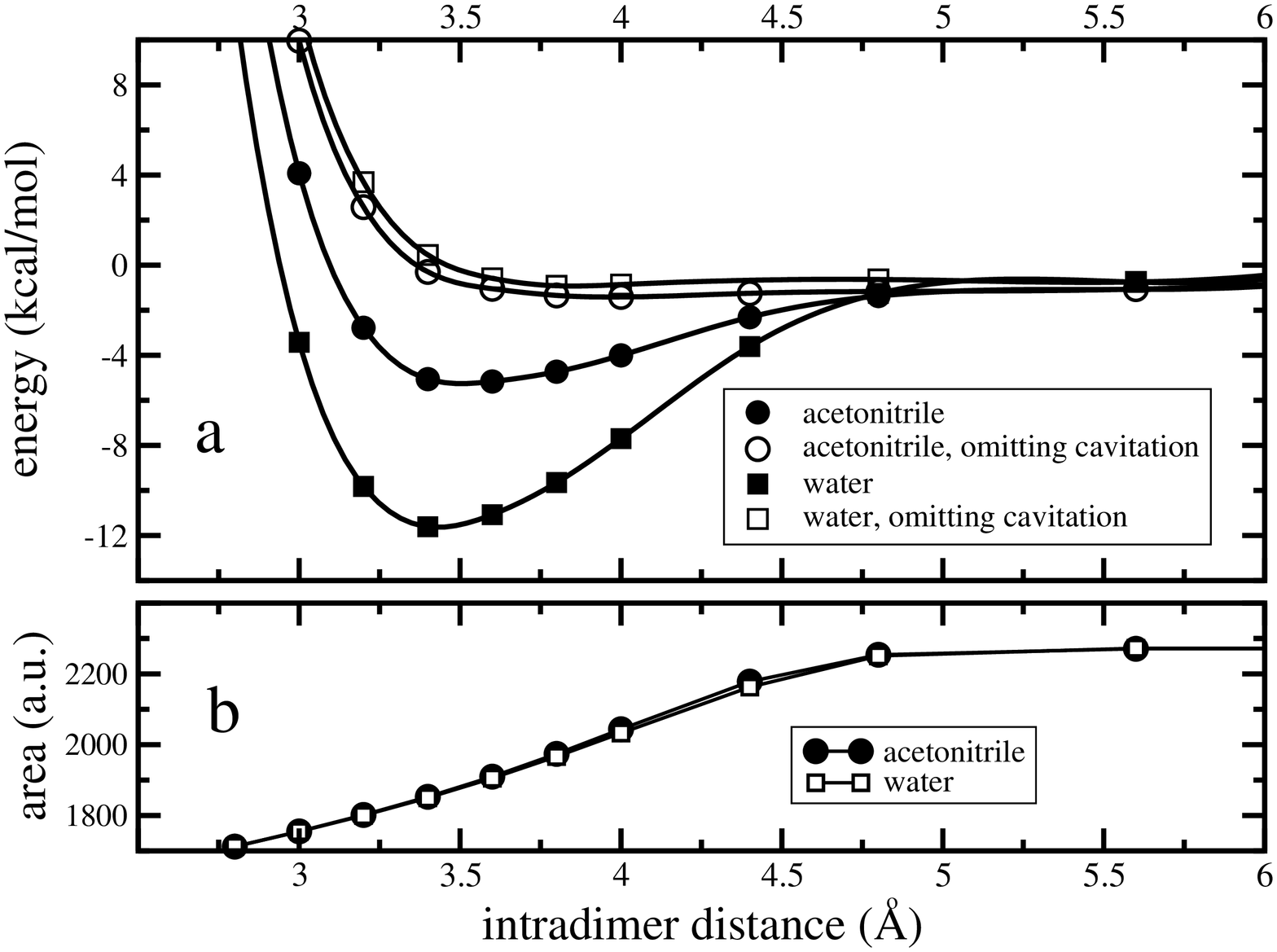}}}
}
\caption{D. Scherlis}
\label{cav}
\end{figure}

\newpage

\begin{figure} \centerline{
\rotatebox{-90}{\resizebox{5in}{!}{\includegraphics{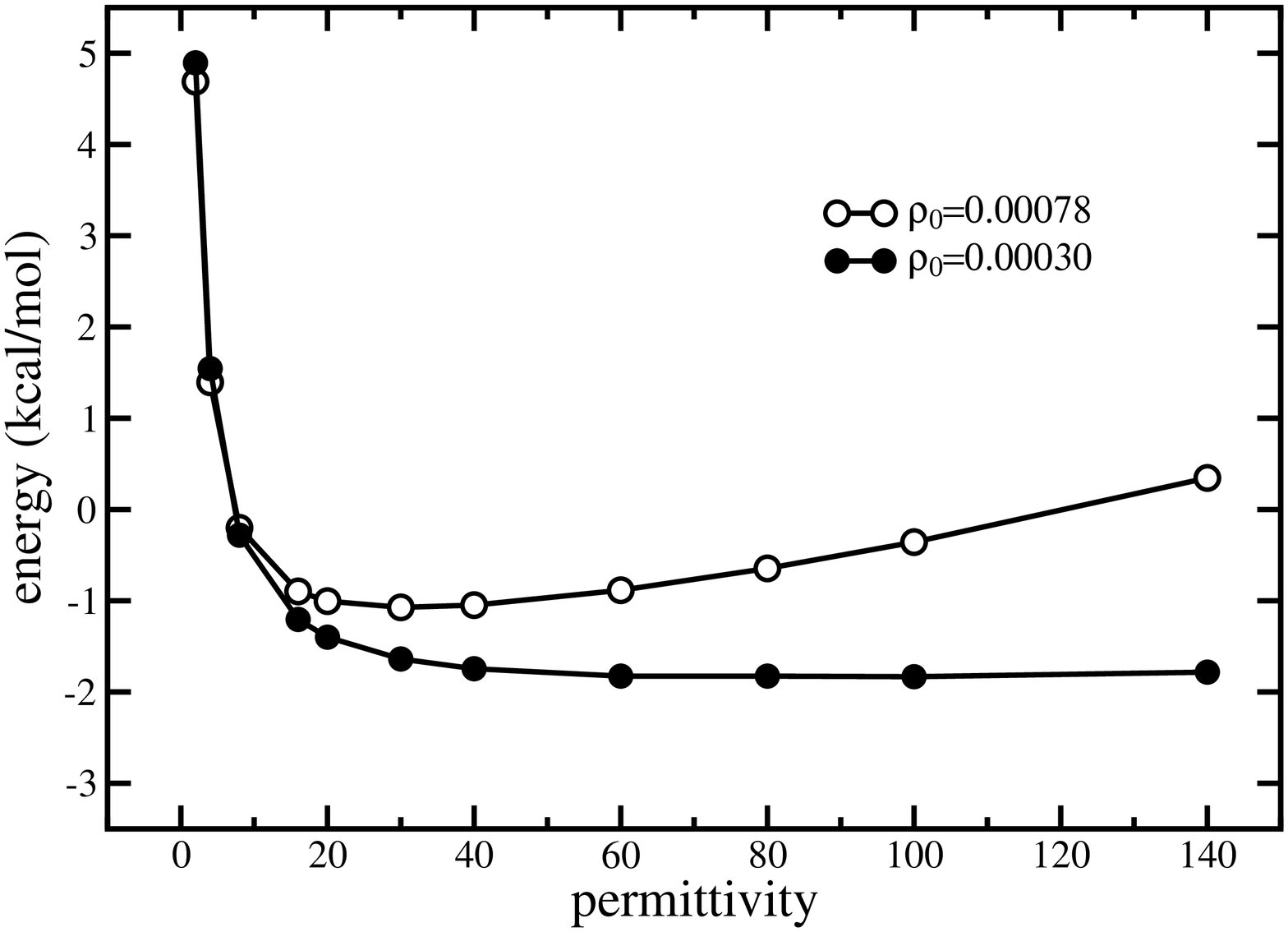}}}
}
\caption{D. Scherlis}
\label{epsilon}
\end{figure}

\newpage

\begin{figure} 
\caption{D. Scherlis}
\label{trimer}
\end{figure}

\newpage

\begin{figure} \centerline{
\rotatebox{-90}{\resizebox{5in}{!}{\includegraphics{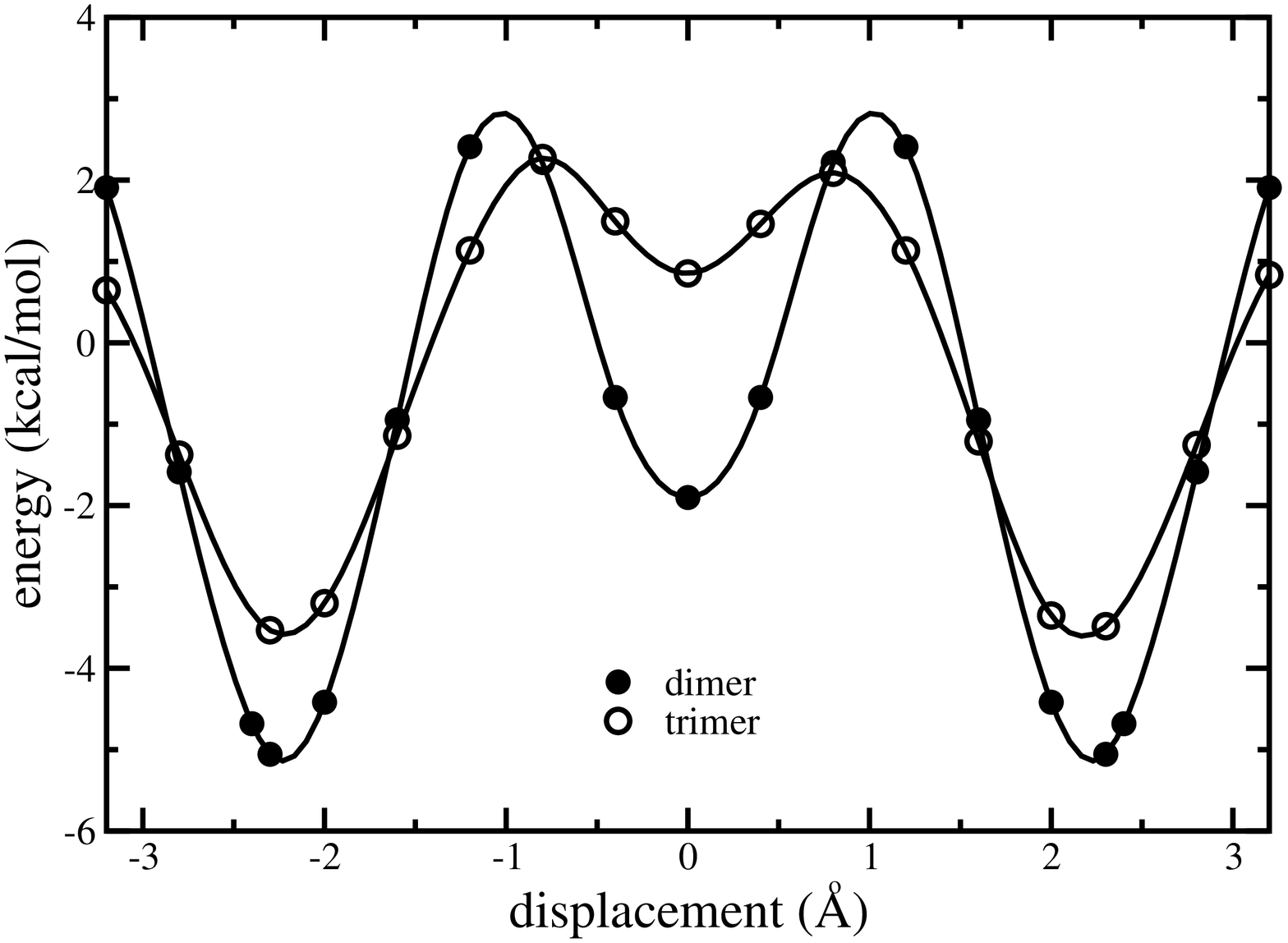}}}
}
\caption{D. Scherlis}
\label{trimerevsl}
\end{figure}

\newpage

\begin{figure} \centerline{
\rotatebox{-90}{\resizebox{5in}{!}{\includegraphics{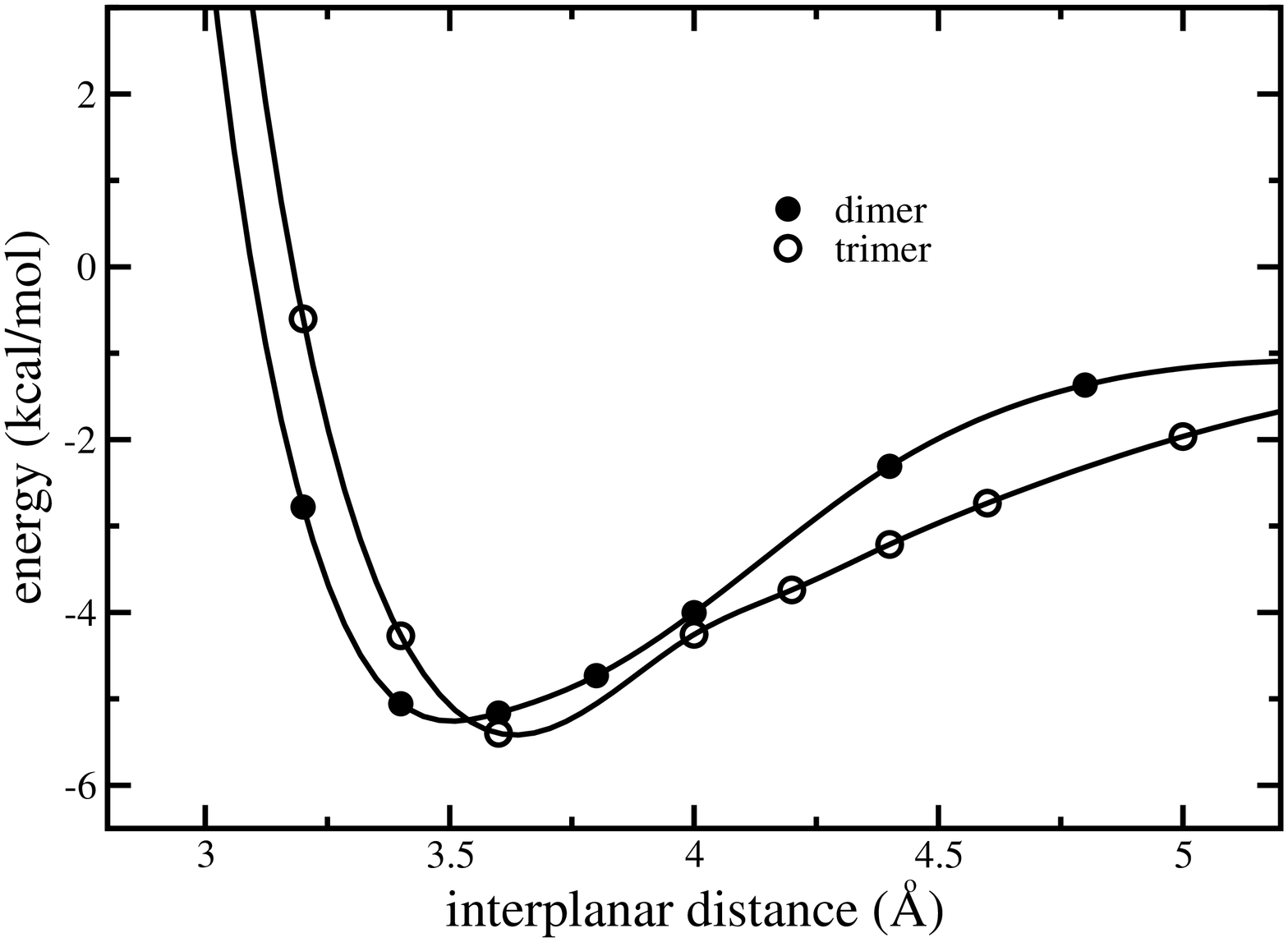}}}
}
\caption{D. Scherlis}
\label{trimerevsd}
\end{figure}


\begin{thebibliography}{10}
\bibitem{nota1}
Standard density-functional theory would probably be a poor choice 
to explore this trend in long oxidized oligomers, given its
inability to represent charge localization
in a long, linear $\pi$-conjugated system, consequence of the
self-interaction error. Alternative schemes as
post-Hartree-Fock methods would be more suited to treat this problem.
\end{thebibliography}
\end{document}